\documentclass[aps,prl,twocolumn]{revtex4}

\usepackage{graphicx}
\usepackage{amsmath}
\usepackage{amssymb}
\usepackage{dcolumn}

\begin{document}

\newcommand{\dmskcomment}[1]{\marginpar{$\spadesuit$}}
\newcommand{\ftot}{{F_{\mbox{\scriptsize tot}}}}
\newcommand{\gammamax}{\gamma_{\mbox{\scriptsize max}}}

\title{Amplification of Fluctuations in a Spinor Bose Einstein
Condensate}

\author{S. R. Leslie$^1$}
\email{sleslie@berkeley.edu}
 \author{ J. Guzman$^{1,2}$, M. Vengalattore$^1$, J. D. Sau$^1$, M. L. Cohen$^{1,2}$, D. M. Stamper-Kurn$^{1,2}$}%
 \affiliation{
    $^1$Department of Physics, University of California, Berkeley CA 94720 \\
    $^2$Materials Sciences Division, Lawrence Berkeley National Laboratory, Berkeley, CA 94720}
\date{\today}%

\begin{abstract}
Dynamical instabilities due to spin-mixing collisions in a $^{87}$Rb
$F=1$ spinor Bose-Einstein condensate are used as an amplifier of
quantum spin fluctuations. We demonstrate the spectrum of this
amplifier to be tunable, in quantitative agreement with mean-field
calculations. We quantify the microscopic spin fluctuations of the
initially paramagnetic condensate by applying this amplifier and
measuring the resulting macroscopic magnetization. The magnitude of
these fluctuations is consistent with predictions of a
beyond-mean-field theory. The spinor-condensate-based spin amplifier
is thus shown to be nearly quantum-limited at a gain as high as 30
dB.
\end{abstract}

\maketitle

Accompanied by a precise theoretical framework and created in the
lab in a highly controlled manner, ultracold atomic systems serve as
a platform for studies of quantum dynamics and many-body quantum
phases.  Among these systems, gaseous spinor Bose Einstein
condensates \cite{ho98,ohmi98,sten98spin,schm04,chan04}, in which
atoms may explore all sub-levels of a non-zero hyperfine spin $F$,
provide a compelling opportunity to access the static and dynamical
properties of a magnetic superfluid
\cite{mies99meta,stam99tunprl,chan05nphys,sadl06symm,veng08helix}.

We previously identified a quantum phase transition in an $F=1$
spinor Bose Einstein condensate between a paramagnetic and
ferromagnetic phase \cite{sadl06symm}.  This transition is crossed
as the quadratic Zeeman energy term, of the form $q F_z^2$, is tuned
through a critical value $q = q_0$; here, $F_z$ is the longitudinal
($\hat{z}$ axis) projection of the dimensionless vector spin
operator $\bf{F}$. Accompanying this phase transition is the onset
of a dynamical instability in a condensate prepared in the
paramagnetic ground state, with macroscopic occupation of the $|m_z=
0\rangle$ magnetic sublevel
\cite{lama07quench,sait07quench,zhan05instab}. This instability
causes transverse spin perturbations to grow exponentially,
producing atoms into the $|m_z = \pm 1\rangle$ sublevels.  In
contradiction with the mean-field prediction that the paramagnetic
state should remain stationary because it lacks fluctuations by
which to seed the instability, experiments revealed the spontaneous
magnetization of such condensates after they were rapidly quenched
across the phase transition.

In this Letter, we investigate the spin fluctuations that become
amplified by the spin-mixing instability. In particular, we test
whether these fluctuations correspond to quantum noise, i.e.\ to the
zero-point fluctuations of quantized spin excitation modes that
become unstable.  For this, we use the spin-mixing instability as an
amplifier, evolving microscopic quantum fluctuations into measurable
macroscopic magnetization patterns. We present two main results.
First, we characterize the spin-mixing amplifier and demonstrate its
spectrum to be tunable by varying the quadratic Zeeman shift. This
spectrum compares well with a theoretical model that accounts for
the inhomogeneous condensate density and for magnetic dipole
interactions. Second, we measure precisely the transverse
magnetization produced by this amplifier at various stages of
amplification, up to a gain of 30 dB in the magnetization variance.
This magnetization signal corresponds to the amplification of
initial fluctuations with a variance slightly greater than that
expected for zero-point fluctuations.

Descriptions of the dynamics of initially paramagnetic spinor
condensates
\cite{lama07quench,dams07quench,uhlm07quench,sait07quench,sait05spont,mias08}
have focused on the effects of the quadratic Zeeman shift and of the
spin-dependent contact interaction.  The latter interaction has the
mean-field energy density $c_2 n \langle \bf{F} \rangle^2$, and,
with $c_2 = 4 \pi \hbar^2 \Delta a / 3m < 0$, favors a ferromagnetic
state; here, $\Delta a=(a_{2}-a_{0})$ where $a_\ftot$ is the
$s$-wave scattering length for collisions between particles of total
spin $\ftot$ \cite{ho98,ohmi98} and $m$ is the atomic mass.
Excitations of the uniform condensate include both the scalar
density excitations and also two polarizations of spin excitations
with a dispersion relation given as
$E_s^2({\bf{k}})=(\varepsilon_k+q)(\varepsilon_k+q-q_0)$, where
$\varepsilon_k=\hbar^2 k^2/2 m$ and $q_0 = 2 c_2 n$. For $q>q_0$,
spin excitations are gapped ($E_s^2
> 0$), and the paramagnetic condensate is stable. Below this
critical value, the paramagnetic phase develops dynamical
instabilities, defined by the condition $E_s^2<0$,  that amplify
transverse magnetization. The dispersion relation defines the
spectrum of this amplification, yielding a wavevector-dependent
time constant for exponential growth of the power in the unstable modes, $\tau = \hbar/2\sqrt{|E_s^2|}$.

The unstable regime is divided further into two regions.  Near the
critical point, reached by a ``shallow'' quench to $q_0/2 \leq q <
q_0$, the fastest-growing instability occurs at zero wavevector,
favoring the ``light-cone'' evolution of magnetization correlations
at ever-longer range \cite{lama07quench}.  For a ``deep'' quench,
with $q<q_0/2$, the instabilities reach a maximum growth rate of
$1/\tau = q_0 / \hbar$. The non-zero wavevector of this
dominant instability sets the size of magnetization domains produced
following the quench.

Experimentally, we characterize the spectrum of this amplifier by
seeding it with broadband noise and then measuring precisely the
spectrum of its output. Similar to previous work \cite{sadl06symm},
we produce condensates of $N_0 = 2.0 \times 10^{6}$ $^{87}$Rb atoms,
with a peak density of $n = 2.6(1) \times 10^{14} \, \mbox{cm}^{-3}$
and a kinetic temperature of $\simeq 50 \, \mbox{nK}$, trapped in a
linearly polarized optical dipole trap characterized by trap
frequencies $(\omega_x,\omega_y,\omega_z)=2\pi\times(39,440,4.2)$
s$^{-1}$. Taking $\Delta a = -1.4(3) \, a_B$ \cite{vankemp02},
with $a_B$ being the Bohr radius, the spin healing length $\xi_s=(8
\pi n |\Delta a|)^{-1/2} = 2.5 \, \mu\mbox{m}$ is larger than the
condensate radius $r_y=1.6\,\mu\mbox{m}$ along the imaging axis
($\hat{y}$) . Thus, the condensate is effectively two-dimensional
with respect to spin dynamics.  For this sample, $q_0 = 2 c_2
\langle n \rangle = h \times 15$ Hz given the maximum $\hat{y}$-axis
column-averaged condensate density $\tilde{n}$.

The quadratic Zeeman shift arises from the application of both
static and modulated magnetic fields.  A constant field of magnitude
$B$, directed along the long axis of the condensate, leads to a
quadratic shift of $q_{B}/h = (70 \, \mbox{Hz}/\mbox{G}^2) B^2$. In
addition, a linearly polarized microwave field, detuned by
$\delta/2\pi = \pm 35 \, \mbox{kHz}$ from the $|F=1, m_z = 0\rangle$
to $|F=2, m_z= 0\rangle$ hyperfine transition, induces a quadratic
(AC) Zeeman shift of $q_{\mu}=- \hbar \Omega^{2}/4\delta$ where
$\Omega$ is the Rabi frequency for the driving field.

The condensate is prepared in the $|m_z = 0 \rangle$ state using rf
pulses followed by application of a $6 \, \mbox{G/cm}$ magnetic
field gradient that expels atoms in the $|m_z = \pm 1\rangle$ states
from the optical trap \cite{sadl06symm}. This preparation takes
place in a static 4 G field and with no microwave irradiation,
setting $q= q_{B} + q_\mu
> q_0$ so that the paramagnetic condensate is stable. Next, we increase the microwave field strength to a constant value, corresponding to a
Rabi frequency in the range of $2 \pi \times (0$ -- $1.5)\,
\mbox{kHz}$, to set $q_\mu$. To initiate
the instability, we ramp the magnetic field over 5
ms to a value of $B = 230 \, \mbox{mG}$ (giving $q_B/h = 7.6 \,
\mbox{Hz}$). During separate repetitions of the experiment (for different values of $q_\mu$),  the
quadratic Zeeman shift at the end of the ramp was thus brought to final
values $q_f/h$ between -2 and 16 Hz.

\begin{figure}[tb]
\centering
\includegraphics[width=0.4\textwidth]{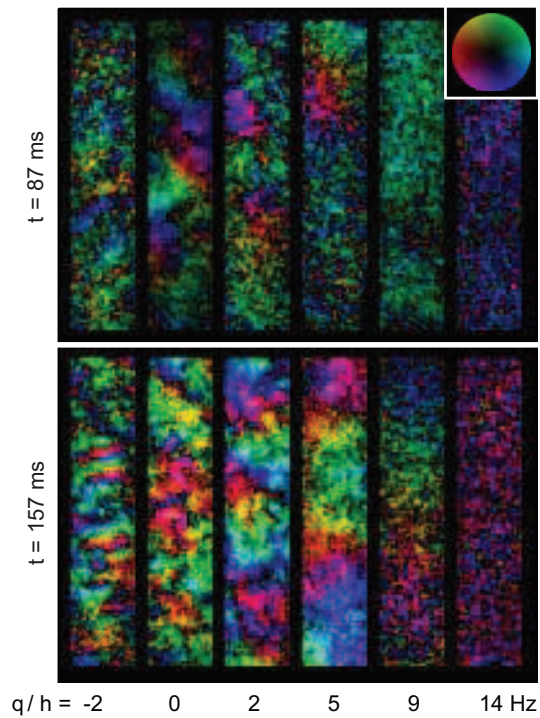}
\caption{Transverse magnetization produced near the condensate
center after 87 ms (top) and 157 ms (bottom) of amplification at
variable quadratic Zeeman shift $q_f$. Magnetization orientation is
indicated by hue and amplitude by brightness (color wheel shown).
The characteristic spin domain size grows as $q_f$ increases.  The
reduced signal strength for $q_f/h \geq 9$ Hz reveals the lower
temporal gain of the spin-mixing amplifier. } \label{fig1}
\end{figure}

Following the quench, the condensate spontaneously develops
macroscopic transverse magnetization, saturating within about 110 ms
to a pattern of spin domains, textures, vortices and domain walls
\cite{sadl06symm}. Using a 2-ms-long sequence of phase-contrast
images, we obtain a detailed map of the column-integrated
magnetization ${\bf{\tilde{M}}}$ at a given time after the quench
\cite{higb05larmor,veng07mag}.  The experiment is then repeated with
a new sample.

The observed transverse magnetization profiles of spinor condensates
(Fig.\ \ref{fig1}) confirm the salient features predicted for the
spin-mixing amplifier \footnote{We measure also the longitudinal
magnetization of the condensate, but, consistent with prior
observations, find that it remains small ($<15\%$ of the maximum
magnetization)}.  The variation of the
amplifier's spatial spectrum with $q_f$ is reflected in the
characteristic size $l_d$ of the observed spin domains, taken as the
distance from the origin at which the magnetization correlation
function,
\begin{equation}
G(\delta\textbf{r})=\frac{\sum_\textbf{r}
\tilde{\textbf{M}}(\textbf{r}+\delta
\textbf{r})\cdot\tilde{\textbf{M}}(\textbf{r})}{(g_F \mu_B)2
\sum_\textbf{r} \tilde{n}(\textbf{r}+\delta
\textbf{r})\tilde{n}(\textbf{r})} , \label{corr}
\end{equation}
acquires its first minimum; here $g_F \mu_B$ is the atomic magnetic
moment. This characteristic size increases with increasing $q_f$ (Fig. 2).
For $q_f / h \geq 9 \, \mbox{Hz}$, the magnetization features become
long ranged and, thus, dominated by residual magnetic field
inhomogeneities ($< 2 \, \mu$G).

\begin{figure}[tb]
\centering
\includegraphics[width=0.4\textwidth]{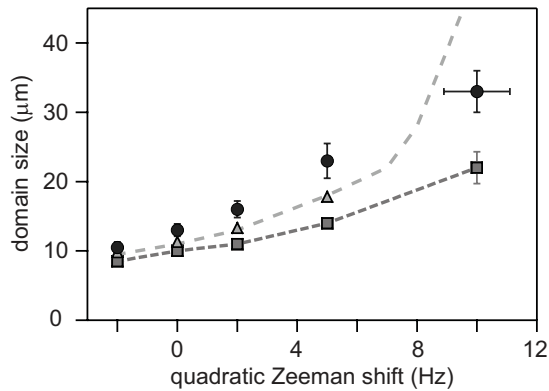}\\
\caption{Characteristic domain size after 87 ms of amplification at
variable quadratic Zeeman shift $q_f$.  Data (circles) are averages
over 5 experimental repetitions; error bars are statistical.
Horizontal error bar reflects systematic uncertainty in $q_f$.
Predictions based on numerical simulations for $|\Delta a| =
1.45\,a_B$ \cite{chan05nphys} (squares) and $1.07\,a_B$
\cite{wide06precision} (triangles) are shown, with error bar
reflecting systematic uncertainty in the atomic
density.}\label{fig2}
\end{figure}

The data also confirm the distinction between deep and shallow
quenches.  The spatially averaged magnetization strength during the
amplification, quantified by $G(0)$ at $t = 87$ ms after the quench,
is found to be constant for $0<q_f/h<6$ Hz, reflecting that the
temporal gain of the amplifier is uniform over $0 \leq q
\leq q_0/2$ \cite{lama07quench}. For shallow quenches, with $q_{f}/h\geq7$ Hz,
the measured magnetization decreases, reflecting a diminishing gain
with increasing $q_f$ up to the transition point.

While the above observations are consistent with theoretical
predictions, we note  the unexpected outcome of quenches to negative
values of $q_f$.  Such quenches revealed a diminished amplifier
gain, resulting in a complete suppression of the growth of
magnetization for $q_f/h\leq -7$ Hz.  We are unable to account for
this behavior.

Having characterized the spin-mixing amplifier, let us consider the
source of its input signal.  For this, we develop a quantum field
description of the spin-mixing instability \cite{lama07quench},
working in the polar spin basis, where $\hat{\phi}_{n,k}$ is the
annihilation operator for atoms of wavevector $k$ in the
zero-eigenvalue states of $F \cdot {\textbf{n}}$. Treating a uniform
condensate within the Bogoliubov approximation, one defines mode
operators $\hat{b}_{n,k} = u \hat{\phi}_{n,k} + v
\hat{\phi}^\dagger_{n,-k}$ for the two polarizations of transverse
($n \in \{x,y\}$) spin excitations.  The spin-dependent many-body
Hamiltonian $H_{s}$ is then approximated as representing two
independent parametric amplifiers:
\begin{equation}
H_{s}= -\frac{i}{2} \sum_k \! |E_s(k)| \! \left(
b_{x,k}^2-b_{x,k}^{\dagger 2}+b_{y,k}^2-b_{y,k}^{\dagger 2}\right).
\end{equation}
The parametric amplifiers serve to squeeze the initial state in each
spin excitation mode, amplifying one quadrature of $b_{n,k}$ (its
real part) and de-amplifying the other.

The above treatment may be recast in terms of spin fluctuations atop
the paramagnetic state: fluctuations of the transverse spin,
represented by the observables $F_x$ and $F_y$, and fluctuations in
the alignment of the spinor, represented by the components $N_{yz}$
and $N_{xz}$ of the spin quadrupole tensor.  We identify the
Bogoliubov operators defined above as linear combinations of these
observables.  Based on this identification, we draw two conclusions.
First, an ideally prepared paramagnetic condensate is characterized
by quantum fluctuations of the Bogoliubov modes.  In the linear
regime, fluctuations in $b_{x,k}$ ($b_{y,k}$) correspond to
projection noise for the conjugate observables $F_{x}$ ($F_y$) and
$N_{yz}$ ($N_{xz}$). Second, the dynamical instabilities lead to a
coherent amplification of these initial shot-noise fluctuations.
While in the present work we observe only the magnetization, in
future work both quadratures of the spin-mixing amplifier  may be
measured using optical probes of the condensate nematicity
\cite{caru04imag} or by using quadratic Zeeman shifts to rotate the
spin quadrature axes.

\begin{figure}[tb]
\centering
\includegraphics[width=0.4\textwidth]{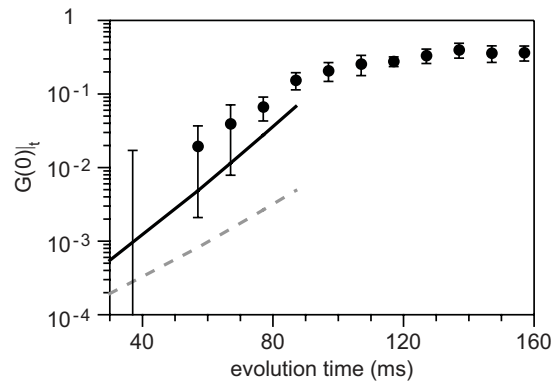}\\
\caption{Temporal evolution of the transverse magnetization variance
$G(0)$, evaluated over the central $16 \times 124$ $\mu$m region of
the condensate and averaging over 8 experimental repetitions; error
bars are statistical. The contribution to $G(0)$ from imaging noise
was subtracted from the data. Predictions from numerical
calculations for $|\Delta a| = 1.45\,a_B$ and $1.07\,a_B$ are shown
as black and gray lines, respectively.}\label{fig3}
\end{figure}

To test the validity of this quantum amplification theory, we
determine $G(0)$ over the central region of the condensate after
different intervals of amplification (Fig.\ \ref{fig3}).   We
consider the linear-amplification theory to be applicable for $t
\leq 90$ ms, and, following Ref.\ \cite{lama07quench}, perform a
least-squares fit to a function of the form
\begin{equation}
\left.G(0)\right|_t=G(0)|_{t_m} \times {\sqrt{t/t_m}} e^{(t-t_m)/\tau}
\end{equation}
Here $\tau$ is the time constant characterizing the growth rate of
the magnetization variance and $G(0)|_{t_m}$ is the value of $G(0)$
at time $t_m=77$ ms.

\begin{figure}[tb]
\centering
\includegraphics[width=0.4\textwidth]{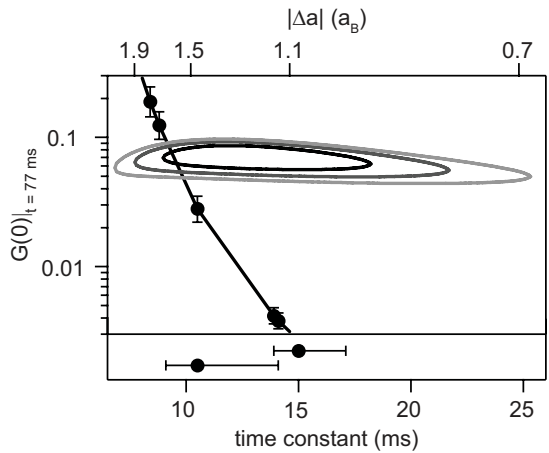}\\
\caption{The magnetization variance $G(0)_{t_m}$ at $t_m = 77$ ms
and exponential time constant $\tau$ of the amplifier, obtained by
fitting data in Fig.\ \ref{fig3}, are indicated by contours of the
1, 2, and 3 $\sigma$ confidence regions using a $\chi^2$ test.
Predictions from numerical calculations of the zero-temperature
quantum amplification theory, assuming different values of $|\Delta
a|$, are shown (circles and interpolating line). Error bars reflect
systematic uncertainty in $q_f$ and in the condensate density.  Time
constants corresponding to reported values for $|\Delta a|$ are
indicated at bottom.}\label{fig4}
\end{figure}

To compare our measurements to the amplifier theory outlined above,
we performed numerical calculations of $\left.G(0)\right|_{t}$,
taking into account the the inhomogeneous density profile; dipolar
interactions; proper position-space, rather than momentum-space,
spin excitation modes; and quantum fluctuations of the initial state
\footnote{J.D.\ Sau et al., to be published}. Such calculations, results of which are shown in
Figs.\ \ref{fig3} and \ref{fig4}, were performed for several values
of the scattering length difference $\Delta a$ within the range of
recent measurements \cite{chan05nphys,wide06precision}.  Our data
are consistent with the quantum-limited amplification of zero-point
quantum fluctuations in the case that $|\Delta a|$ lies in the upper
range of its reported values.  Taking $\tau$ to be determined
instead solely on the basis of our measurements, the magnetization
variance is measured to be between 1 and 50 times greater than that
predicted by the zero-temperature quantum theory.

Several investigations were performed to identify technical or
thermal contributions on top of the expected quantum spin
fluctuations of our samples.  We place an upper bound on thermal
noise by performing a Stern-Gerlach analysis of populations $N_\pm$
in the $|m_z = \pm 1\rangle$ states just after the quench. Obtaining
$N_\pm \leq 3 \times 10^2$ and assuming an incoherent admixture of
Zeeman sublevels, the thermal contribution to $G(0)|_0$ is below $(2
N_\pm / N_0) = 3 \times 10^{-4}$. We confirmed that our results are
insensitive to variations in the gradient strength, duration, and
orientation used during the initial state preparation, and also to
the delay (varied between 0 and 110 ms) between this preparation and
the initiation of the spin amplifier. We checked against technical
noise that would induce extrinsic Zeeman transitions during the
experiment, finding that a condensate starting in the $|m_z =
-1\rangle$ state remained so for evolution times up to 400 ms
following the quench. Altogether, these results suggest that
paramagnetic samples were produced with a near-zero spin
temperature, to which the quantum amplification theory may be
expected to apply.

The microwave fields used to vary the quadratic Zeeman shift cause a
slight mixing between the $F=1$ and $F=2$ hyperfine levels.  We
checked for the influence of this admixture by varying the relative
contributions of the static and modulated field contributions to the
final quadratic Zeeman shift $q_f = q_B + q_\mu$.  As expected given
the small value of $(\Omega / \delta)^2 < 10^{-3}$, no variation in
the magnetization evolution for constant $q_f$ was observed.

It remains uncertain whether the zero-temperature amplifier theory
should remain accurate, out to a gain in the magnetization variance
as high as 30 dB, in a non-zero temperature gas subject to constant
heating and evaporation from the finite-depth optical trap. Indeed,
previous work showed a strong influence of the non-condensed gas on
spin dynamics in a two-component gaseous mixture
\cite{mcgu03normal}.  We compared the amplification of magnetization
at kinetic temperatures of 50 and 85 nK, obtained for different
optical trap depths.  We observed no variation, but note that the
condensate fraction was not substantially varied in this comparison.

In this work, we have demonstrated near-quantum-limited
amplification of magnetization fluctuations using dynamical
instabilities in a spinor Bose gas.  Just as the demonstration of
matter-wave amplification in scalar condensates suggested a host of
applications \cite{deng99fourwave,kozu99amp,inou00amp}, the
spin-mixing amplifier may serve as an important experimental tool in
probing systems with multiple degrees of freedom or in spatially
resolved magnetometry \cite{veng07mag}.

We acknowledge insightful discussions with J.\ Moore and S.\
Mukerjee and experimental assistance from C.\ Smallwood. This work
was supported by the NSF, the David and Lucile Packard Foundation,
DARPA's OLE Program, and the U.S. Department of Energy under Contract No. DE-AC02-05CH11231. Partial personnel and equipment support was
provided by the Division of Materials Sciences and Engineering,
Office of Basic Energy Sciences. S.R.L. acknowledges support from
the NSERC. 

\bibliographystyle{apsrev}

\end{document}